\documentclass{PoS}
\usepackage{siunitx}
\usepackage{subcaption}

\title{The Cherenkov Telescope Array Performance in Divergent Mode}

\ShortTitle{Divergent Pointing Mode for CTA}

\author{\speaker{A. Donini}$^{1,2}$, T. Gasparetto$^{2,3,4}$, J. Bregeon$^{5}$, F. Di Pierro$^{6}$, F. Longo$^{2,3}$, G.~Maier$^{7}$, A.~Moralejo$^{8}$, T. Vuillaume$^{4}$ for the CTA Consortium\footnote{for consortium list see PoS(ICRC2019)1177}\\
        \llap{$^1$}University of Udine, Italy, \hspace{0.1cm} \\
        \llap{$^2$}Istituto Nazionale di Fisica Nucleare (INFN) Trieste, Italy, \hspace{0.1cm} \\
        \llap{$^3$}University of Trieste, Italy, \hspace{0.1cm} \\
        \llap{$^4$}Laboratoire d'Annecy de Physique des Particules, Univ. Grenoble Alpes, Univ. Savoie Mont Blanc, CNRS, LAPP, 74000 Annecy, France,\hspace{0.1cm} \\
        \llap{$^5$}Laboratoire Univers et Particules de Montpellier (LUPM), CNRS-IN2P3, Universit\'e de Montpellier, France,\hspace{0.1cm} \\
        \llap{$^6$}Istituto Nazionale di Fisica Nucleare (INFN) Torino, Italy, \hspace{0.1cm} \\
        \llap{$^7$}Deutsches Elektronen-Synchrotron (DESY), Zeuthen, Germany, \hspace{0.1cm}\\
        \llap{$^8$}Institut de F\'isica d'Altes Energies (IFAE), The Barcelona Institute of Science and Technology, Spain \\
        }

\abstract{Two of the Key Science Projects of the Cherenkov Telescope Array (CTA) consist in performing a deep survey of the Galactic and Extragalactic sky, providing an unbiased view of the Universe at energies above tens of GeV.
To optimize the time spent to perform the Extragalactic survey, a so-called ``divergent mode'' of the CTA was proposed as an alternative observation strategy to the traditional parallel pointing in order to increase its instantaneous field of view.  
The search for transient VHE sources would also benefit from an extended field of view. 
In the divergent mode, each telescope points to a position in the sky that is slightly offset, in the outward direction, from the center of the field of view.
In this contribution, we present the first performance estimation from full Monte Carlo simulation of possible CTA divergent mode setups.
}

\FullConference{36th International Cosmic Ray Conference -ICRC2019-\\
		July 24th - August 1st, 2019\\
		Madison, WI, U.S.A.}

\begin{document}

\section{Introduction}
    
    The Cherenkov Telescope Array (CTA) is going to be the major next-generation observatory for ground-based very-high-energy gamma-ray astronomy, with a 5 to 20 times better sensitivity than existing similar experiments, thanks to a large number of telescopes (more than 100) built across two sites, one in the Northern Hemisphere, at La Palma in the Canary Islands, and one in the Southern Hemisphere, at Cerro Paranal in Chile.

    The telescopes that are going to compose CTA will have three different sizes: the Small-Sized Telescopes (SSTs) will have a primary mirror of about 4 meters in diameter, the Medium-Sized Telescope (MSTs) with a primary mirror of 12 meters and the Large-Sized Telescopes (LSTs) that will be the largest one, with a 23 meters primary mirror. Those three telescopes sizes are needed in order to cover the energy range between few tens of GeV to hundreds of TeV with a significant improvement in angular resolution, energy resolution and sensitivity with respect to existing IACT experiments (see \cite{cta} and \cite{cta-perf}). 
    
    In this work, we will present the Monte Carlo (MC) simulations carried out in order to study the performances of different divergent pointing configurations in the reconstruction of the direction of primary gamma-rays. We will show some angular resolution plots with different telescope multiplicities.
    A deeper study of other performance parameters (e.g. energy reconstruction or background suppression capabilities) will not be present here, but they are to a large extent dependent on the stereoscopic reconstruction and they will be taken into account in the future.
    

\section{Divergent pointing}
    
    The Galactic and extragalactic surveys are two of the main proposed legacy projects of CTA, providing an unbiased view of the Universe at energies above tens of GeV. Considering the limited field of view of the Cherenkov telescopes, the time needed for those science projects is large and since imaging Cherenkov facilities have a limited duty cycle there is a strong motivation in trying to reduce the observation time needed to achieve them.
  
    Surveys  would, in general, be conducted in a mode with telescopes co-pointed in the ``parallel mode'', i.e all telescopes point to the same position in the sky, but the huge number of telescopes of CTA with respect to existing instruments will allow to take full advantage of new pointing modes.

    The performance of an array of telescopes operating in sky-survey mode depends upon the field of view (FoV) of the system and the time of observation needed to achieve a given signal significance level, i.e. its sensitivity. In the standard, conservative, pointing scheme, sky surveys may be performed in parallel mode, however, in such a case the FoV of the array is limited by the FoVs of individual telescopes.
    
    The overall FoV of a telescope array can be significantly enlarged by choosing the pointing direction of each telescope according to its position in the array. In the divergent mode, telescopes point in the outward direction by an angle increasing with the telescope distance from the array center.
    In the determination of the telescope pointing directions, there is a trade-off between total FoV and average telescope multiplicity (number of independent views obtained of a given shower), the latter being a crucial parameter for the quality of the shower reconstruction.
    The chosen telescope pointings will affect angular resolution, energy resolution and sensitivity, that have to be studied in order to decide on the optimal divergent mode parameters.

    A divergent mode is possible and under consideration for the extragalactic survey, offering increased instantaneous FoV and survey depth at the expense of average ``instantaneous'' sensitivity across the FoV and the angular and energy resolutions (more on the surveys with CTA in \cite{surveys}).
    
    The serendipitous detection of transient sources would also benefit from an observation which is being carried with an extended field of view since it may increase the probability of a detection. Moreover, the electromagnetic follow-up of transient events such as neutrinos, gravitational waves and gamma-ray bursts with an increased field of view would allow reducing the time needed to observe a certain patch of sky with a certain sensitivity (see Chapter 8 and 9 of \cite{science-cta} for more connections between science goals and divergent pointing).
    
    A preliminary study of this pointing mode based on the first CTA Monte Carlo productions using only the MSTs has been presented in the past (see~\cite{Lucie}~and~\cite{polacchi}), where the authors have shown that the divergent mode could be superior with respect to the normal pointing mode for source detection, i.e. it will have a superior flux sensitivity.

    Even though, as expected, the angular and energy resolutions for the divergent pointing mode is up to a factor of about two worse when compared to the normal pointing, an increase in flux sensitivity is very attractive for the survey science project. 

    Due to the purpose of this task, the divergent mode favours the use of an array of telescopes with an already large enough field of view per telescope. This is the reason why initially only the subarray composed by the 15 MSTs that will be built in La Palma, without considering the 4 LTSs, was studied. For the participation of LSTs to the extragalactic survey, see section 8.4.4 from~\cite{science-cta}.
    
    This is just the first step of the work since in the southern CTA site there will be more MST to consider. A study of the divergent mode of the SSTs, having a larger field of view with respect to the MSTs and coming in a greater number, will be particularly interesting according to the science case. 

\section{Simulations and data analysis}
    
    \subsection{Monte Carlo production - CORSIKA and sim\_telarray}
    
        In order to perform the simulation of the atmospheric showers, we used CORSIKA (version 6.990) \cite{CORSIKA_code}, which performs a detailed simulation of extensive air-shower that are initiated by high energy primary particles hitting the top of the atmosphere.
        
        
        We simulated gamma-rays from a point-like source coming from an azimuth of 180 deg, corresponding to the geomagnetic South, and at an altitude of 70 degrees above the horizon. The corresponding parameters for the underlying array, such as telescopes positions, were taken from the third massive Monte Carlo production (the \texttt{prod3}) which was done for the La Palma site with 4 LSTs and 15 MSTs. The only difference with respect to the \texttt{prod3} is the pointing of each telescope.
        For each tested configuration we simulated $10^8$ point source gamma-rays, in the energy range \SI{3.0}{\GeV} - \SI{330}{\TeV}.

        The response of the telescope array is then simulated with the sim\_telarray program (see~\cite{sim} and \cite{grid_COR}): while in parallel pointing mode, the pointing of the telescopes is usually unique between all of them, in the case of divergent mode the pointing direction of each telescope had to be manually set by changing the configuration file which is given as input to sim\_telarray.
        
\begin{figure}[t!]
			\begin{subfigure}{0.5\textwidth}
			    \centering
				\includegraphics[width=1.05\linewidth, keepaspectratio]{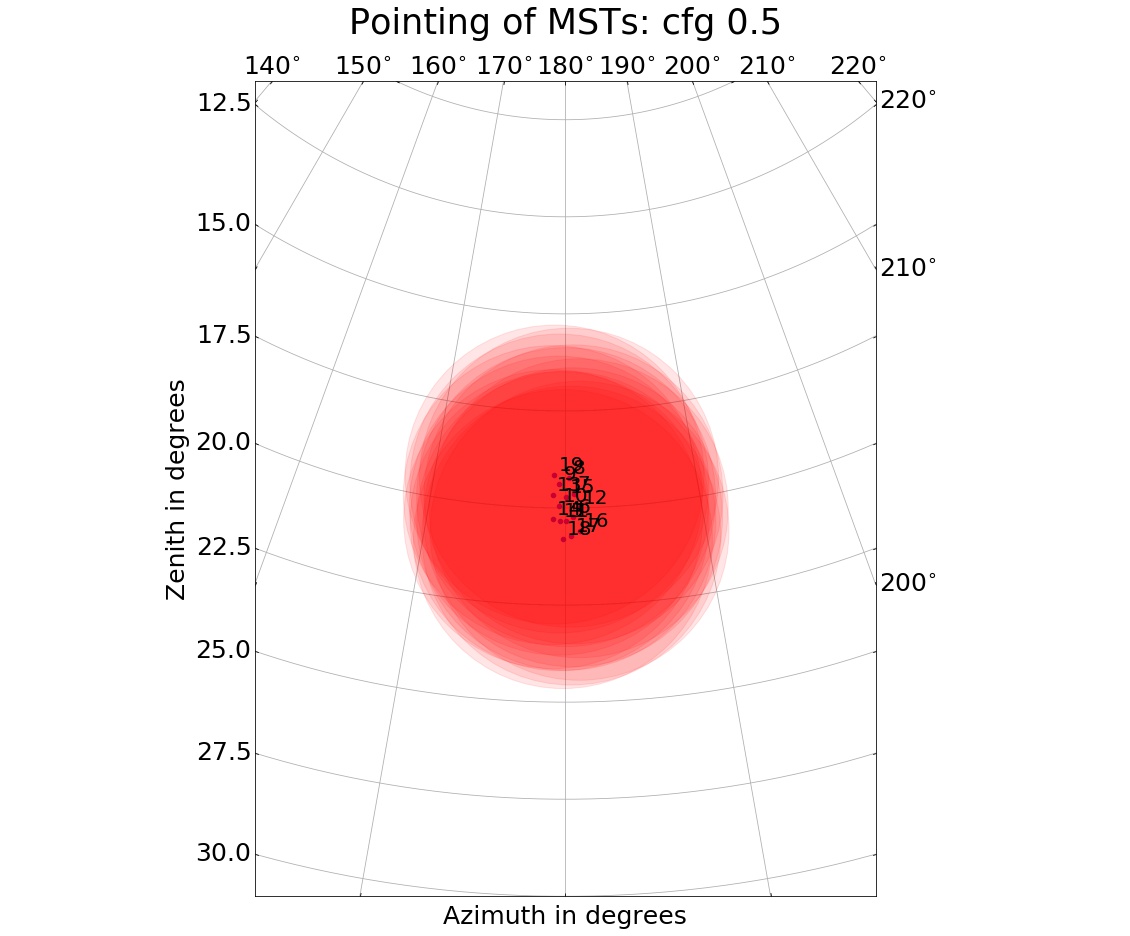}
			    \caption{Offset 0.5.}
            \end{subfigure}
            \hspace{1mm}
            \begin{subfigure}{0.5\textwidth}
			    \centering
    			\includegraphics[width=1.05\linewidth, keepaspectratio]{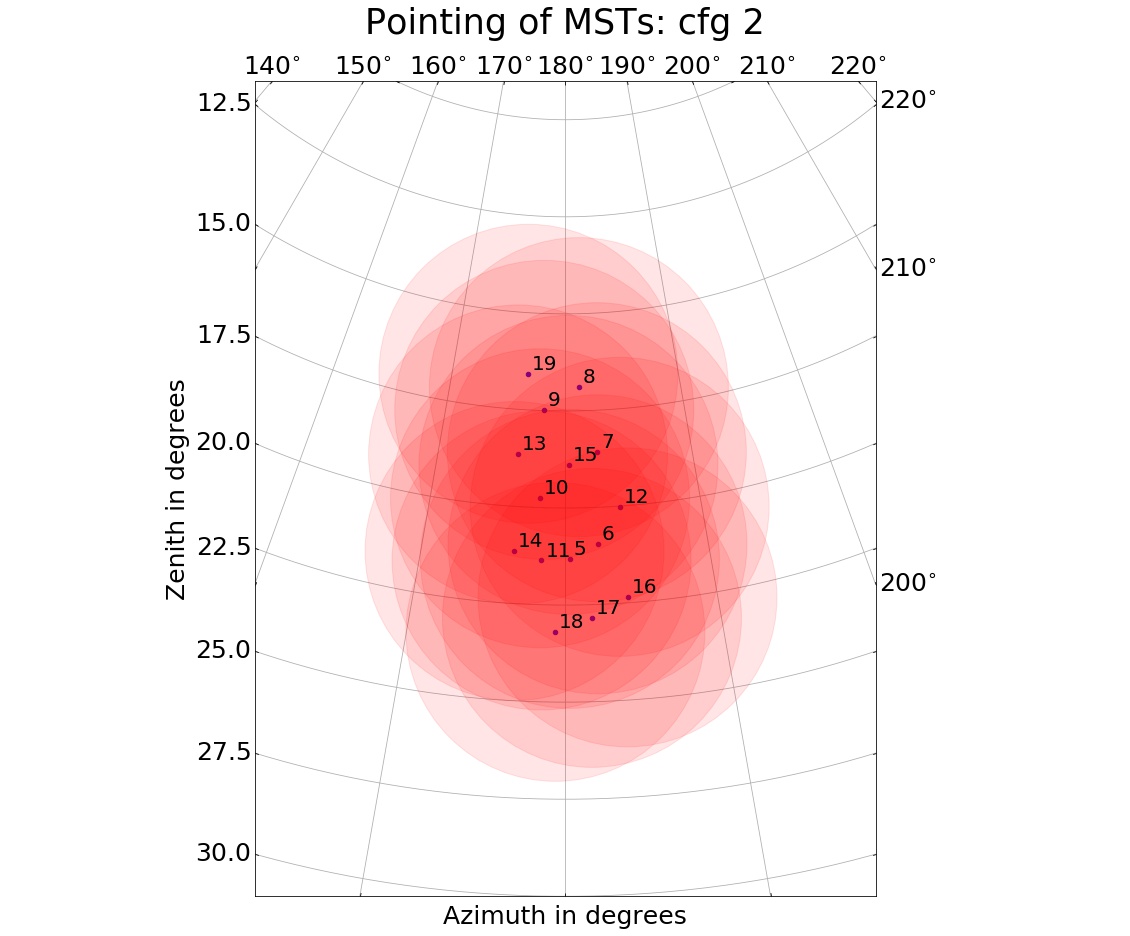}
			    \caption{Offset 2.}
            \end{subfigure}
            \newline
            \begin{subfigure}{0.5\textwidth}
			    \centering
				\includegraphics[width=1.05\linewidth, keepaspectratio]{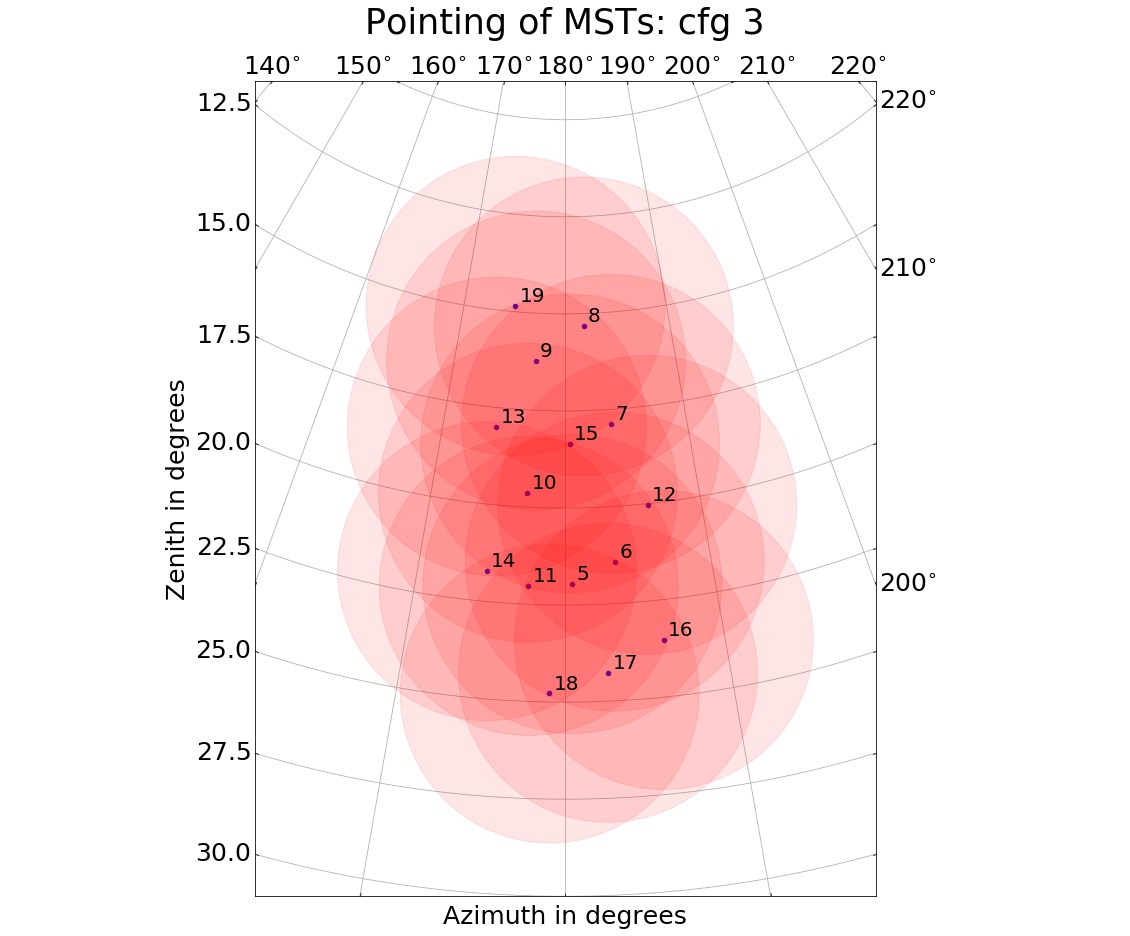}
			    \caption{Offset 3.}
            \end{subfigure}
            \hspace{1mm}
            \begin{subfigure}{0.5\textwidth}
			    \centering
				\includegraphics[width=1.05\linewidth, keepaspectratio]{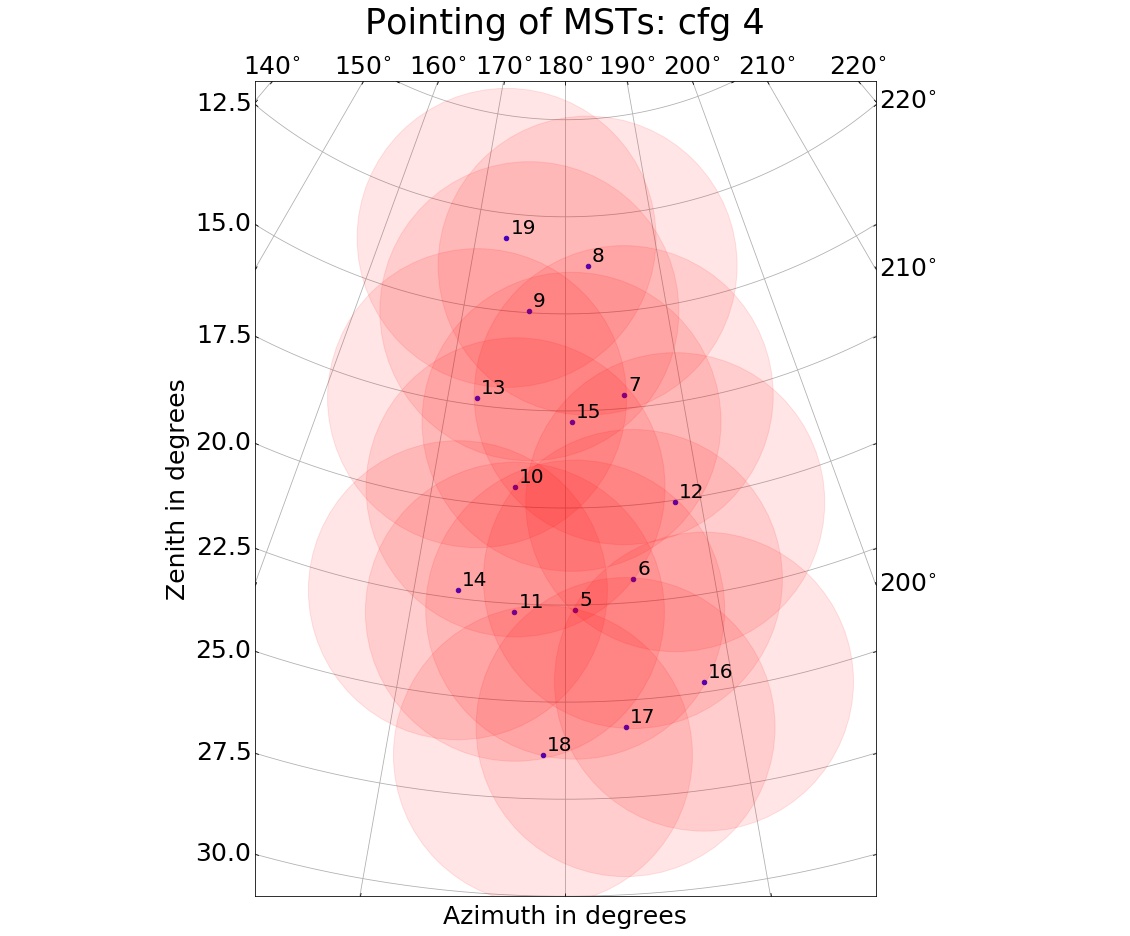}
			    \caption{Offset 4.}
            \end{subfigure}
        \caption{
        Fields of view in polar coordinates for the different ``on-axis'' configurations that have been tested. The camera used for the MSTs is the NectarCam~\cite{NectarCam}: the IDs of the MSTs range from 5 to 19. The ``offset'' is just a parameter used in our script to generate those pointings, and not the actual separation in degrees between the telescopes. The incoming direction of the simulated gamma-rays is 180 deg of zenith and 20 deg of azimuth.
        }
        \label{Fig:fovs}
        \end{figure}

        The pointing was obtained thanks to a separate tool, whose final purpose will be to provide for a given layout and direction of the general pointing center, the best individual telescope azimuth, and zenith angle and a multiplicity map. In figure \ref{Fig:fovs} are reported the four configurations that were simulated, each one with a different offset between the telescope pointing directions.
        
    \subsection{Ctapipe analysis}
        The reconstruction of the simulated data has been done using ctapipe~\cite{ctapipe}, a framework for the data processing for CTA data which is being developed by the members of the CTA Consortium. Ctapipe is mainly developed using Python and its libraries for scientific computing. 

        The simulated events are loaded and the waveforms for the pixels in the cameras are calibrated with the pedestal value: the time slices are then integrated in order to get the charge per pixel belonging to the camera of each telescope triggered in the event. The final images are cleaned with a two-threshold tailcut method and the moments, called Hillas parameters (from~\cite{hillas}), of the resulting elliptical images, are calculated. 
        Given the positions of the telescopes and the parametrization of the ellipse in the camera, a stereo reconstruction can be performed in order to find the impact point of the shower on the ground, its direction and the height at which the Cherenkov signal is maximum~(more on ctapipe in~\cite{ctapipe} and in the code repository).
        
        The direction reconstruction in divergent pointing is a crucial step in the overall analysis and it is the biggest difference with respect to the standard analysis in parallel mode. Two methods for event reconstruction are implemented in ctapipe and we adapted for divergent pointing the one performing better in parallel mode.

    \subsection{Event reconstruction}
   
        The reconstruction method that we have used in the analyses in ctapipe builds, from the image recorded by each telescope, a plane defined by the projection of the shower axis on the camera and the telescope position, placing this in a common 3D reference frame. 
        Those planes are then intersected pair-wise, and the angle between them is used as a weight for the computation of the final reconstructed direction, which is a weighed average between all pair-wise directions. The reconstruction of the direction in the sky did not need any correction with respect to the parallel pointing case.
        The planes were used for the reconstruction of the height of the maximum of the shower, which also did not need any correction for our analyses.
        
        For what concerns the reconstruction of the impact point on the ground, this had to be corrected for the divergent pointing since it is not done in 3D but it is performed in a common plane, called ``TiltedGroundFrame'', which is perpendicular to the array pointing direction. While in parallel mode, the pointing of the array is the same as the telescope pointing, and the camera planes are parallel with respect to the common plane, this is not true any more in divergent pointing: this implies that the angles that are measured in the cameras in divergent mode cannot be reported on the common plane and used for the impact point reconstruction, since angles are not preserved between non-parallel planes. 
        
        In order to properly correct the angle of the ellipse measured in the camera using ctapipe functions, the camera points used to build the 3D plane for each telescope, that are projected as points in the sky, have been re-projected on a ``fake parallel-pointing'' telescope and the tilt angle of the ellipse has been re-calculated in the camera of this fake telescope: all the corrected tilt angles in the cameras have been used for the impact point reconstruction, showing an improvement in its reconstruction. 

\section{Results and discussion}
    \begin{figure}[t!]
		\begin{subfigure}{0.5\textwidth}
		    \centering
			\includegraphics[width=0.9\linewidth, keepaspectratio]{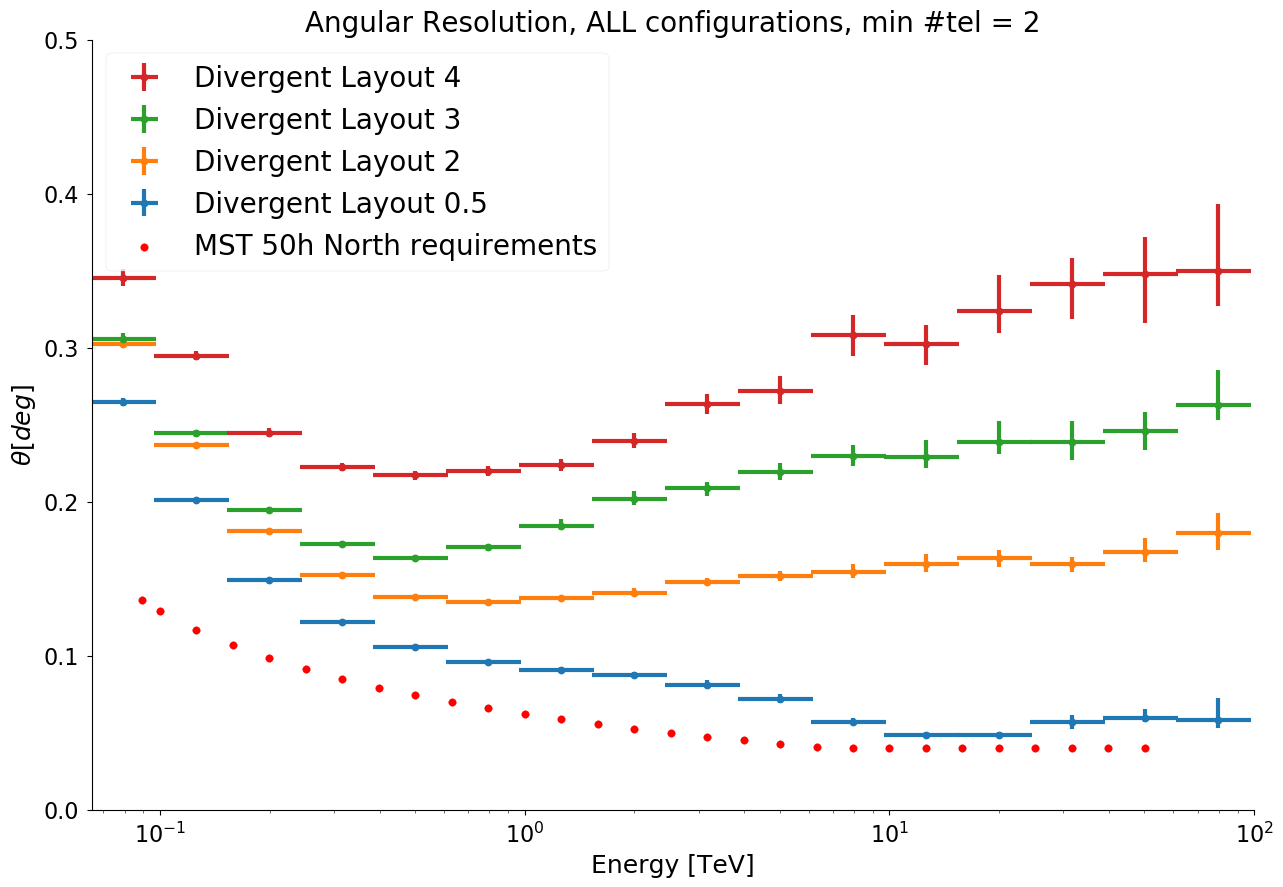}
		    \caption{Minimum multiplicity equal to 2.}
        \end{subfigure}
        \begin{subfigure}{0.5\textwidth}
		    \centering
			\includegraphics[width=0.9\linewidth, keepaspectratio]{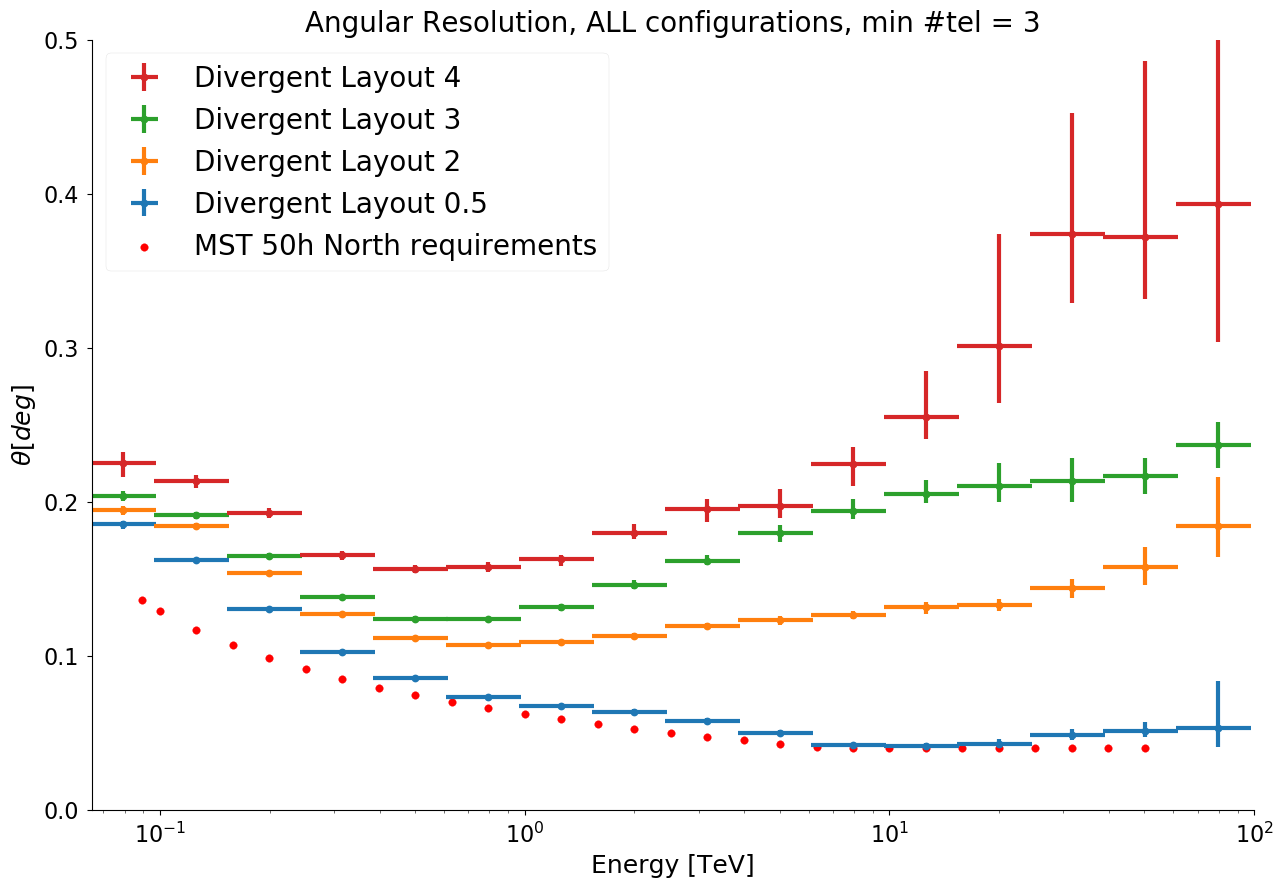}
		    \caption{Minimum multiplicity equal to 3.}
        \end{subfigure}
        \begin{subfigure}{\textwidth}
            \vspace{3mm}
            \centering
			\includegraphics[width=0.5\linewidth, keepaspectratio]{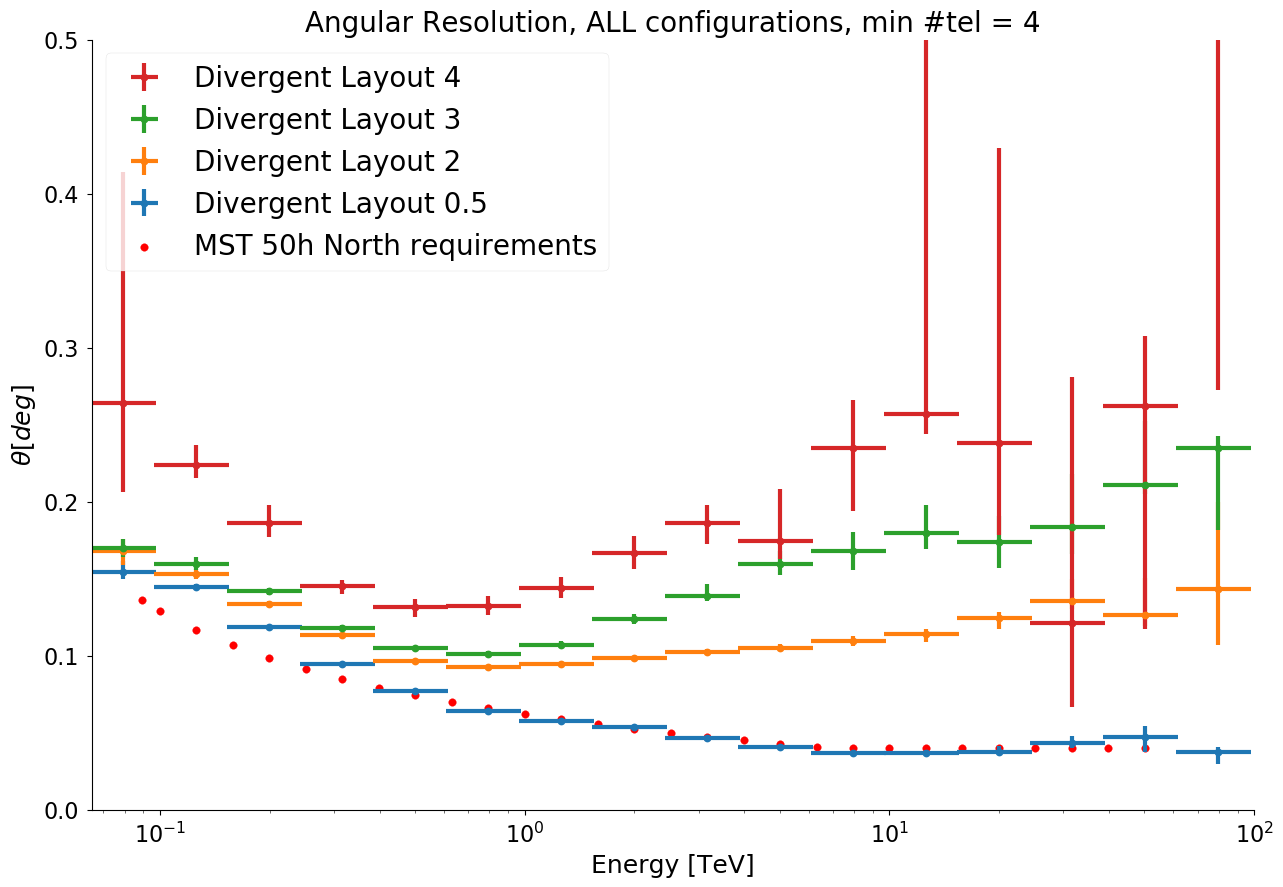}
		    \caption{Minimum multiplicity equal to 4.}
        \end{subfigure}
    \caption{Angular resolution for the ``on-axis'' configuration of the point gamma analysed plotted against the true energy. The plots (a), (b) and (c) have been created selecting only events reconstructed with at least 2, 3 and 4 telescopes triggered. Maximum leakage set to 10\%.}
    \label{Fig:AngResOnAxis}
    \end{figure}
    
    The plots\footnote{All the angular resolution plots have been created with \href{https://github.com/vuillaut/ctaplot}{ctaplot}.} in figure~\ref{Fig:AngResOnAxis} represent the angular resolution that we have obtained for the four configurations that we have been investigating (pointings are in figure~\ref{Fig:fovs}). The plots represent the separation of the reconstructed position in the sky from the true position of the source, with the error bar representing 68\% of the events in that energy bin.

    The angular resolution plots usually represent the angle within which 68\% of reconstructed gamma rays fall, relative to their true direction and are plotted against the reconstructed energy~\cite{cta-perf}: gamma-hadron separation cuts are usually applied to the MC that is being used to determine the angular resolution. In our analyses, we have not taken into account the protons but only point gamma-rays, so no gamma-hadron separation have been carried out and the angular resolution is plotted against the true energy from the simulation.
    
    The plots in figure~\ref{Fig:AngResOnAxis} were obtained applying some selection cuts: the resulting image after the cleaning must have at least 6 pixels left and the leakage, which is the percentage of signal deposited in the border of the camera, had to be lower than 10\%. 
    
    The results that we have obtained show that a more divergent configuration gives a worse angular resolution with respect to a less divergent one, especially at high energies and with low multiplicity. A higher multiplicity gives a better angular resolution over the whole energy range, at the expense of a lower number of events; the configuration labeled ``0.5'', not that different from a parallel pointing configuration, has an angular resolution close to the requirements, as expected, while there is a small difference between the layouts ``2'' and ``3'' below 1 TeV.
    
    We are going to run some analyses with harder cuts on the leakage in order to properly remove the images that fall at the borders of the cameras, therefore leading to a bad reconstruction. 

    These are just preliminary results and more tests and optimizations are necessary to understand the connection between the pointing and the reconstruction.
    
        \begin{figure}[t!]
		\begin{subfigure}{0.5\textwidth}
		    \centering
			\includegraphics[width=0.9\linewidth, keepaspectratio]{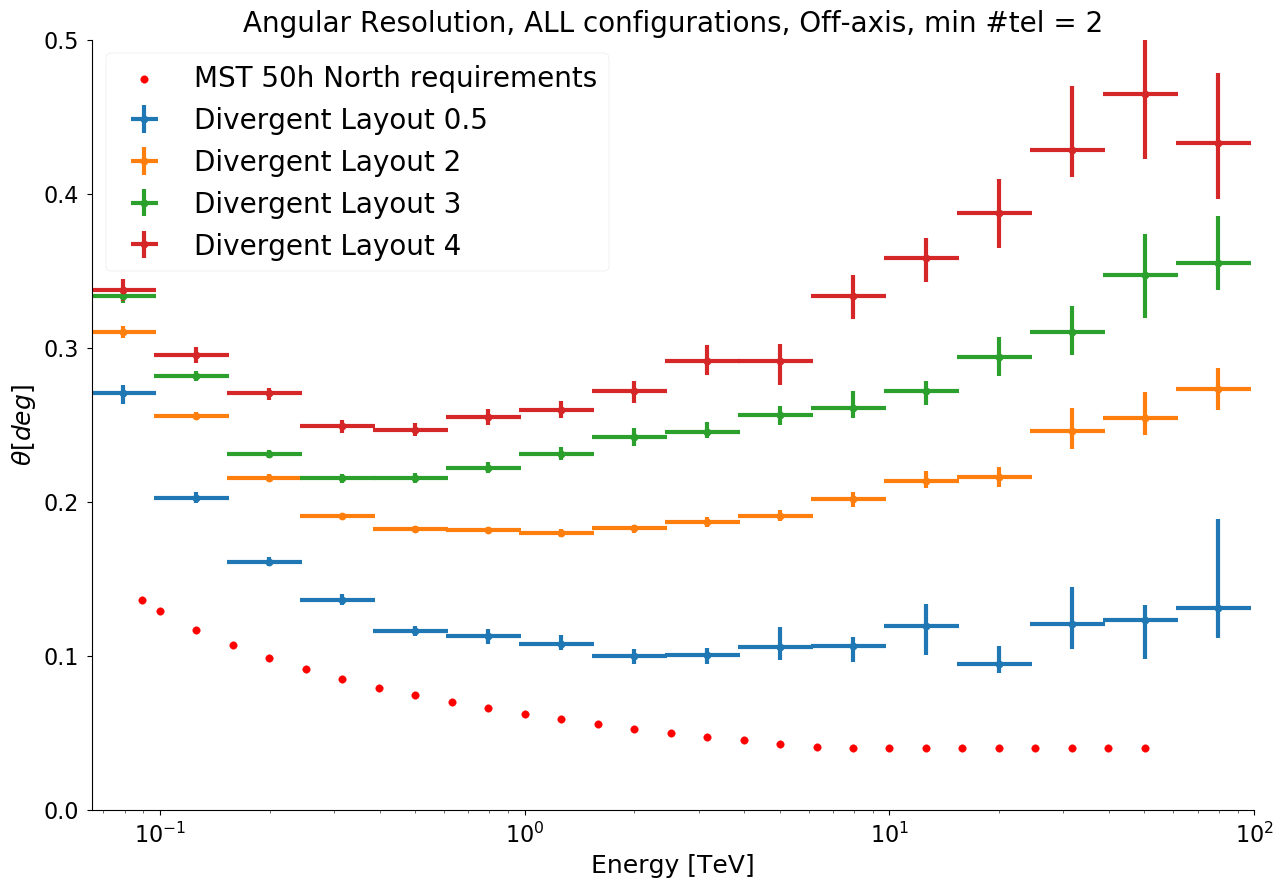}
		    \caption{Minimum multiplicity equal to 2.}
        \end{subfigure}
        \hspace{1mm}
        \begin{subfigure}{0.5\textwidth}
			\includegraphics[width=0.9\linewidth, keepaspectratio]{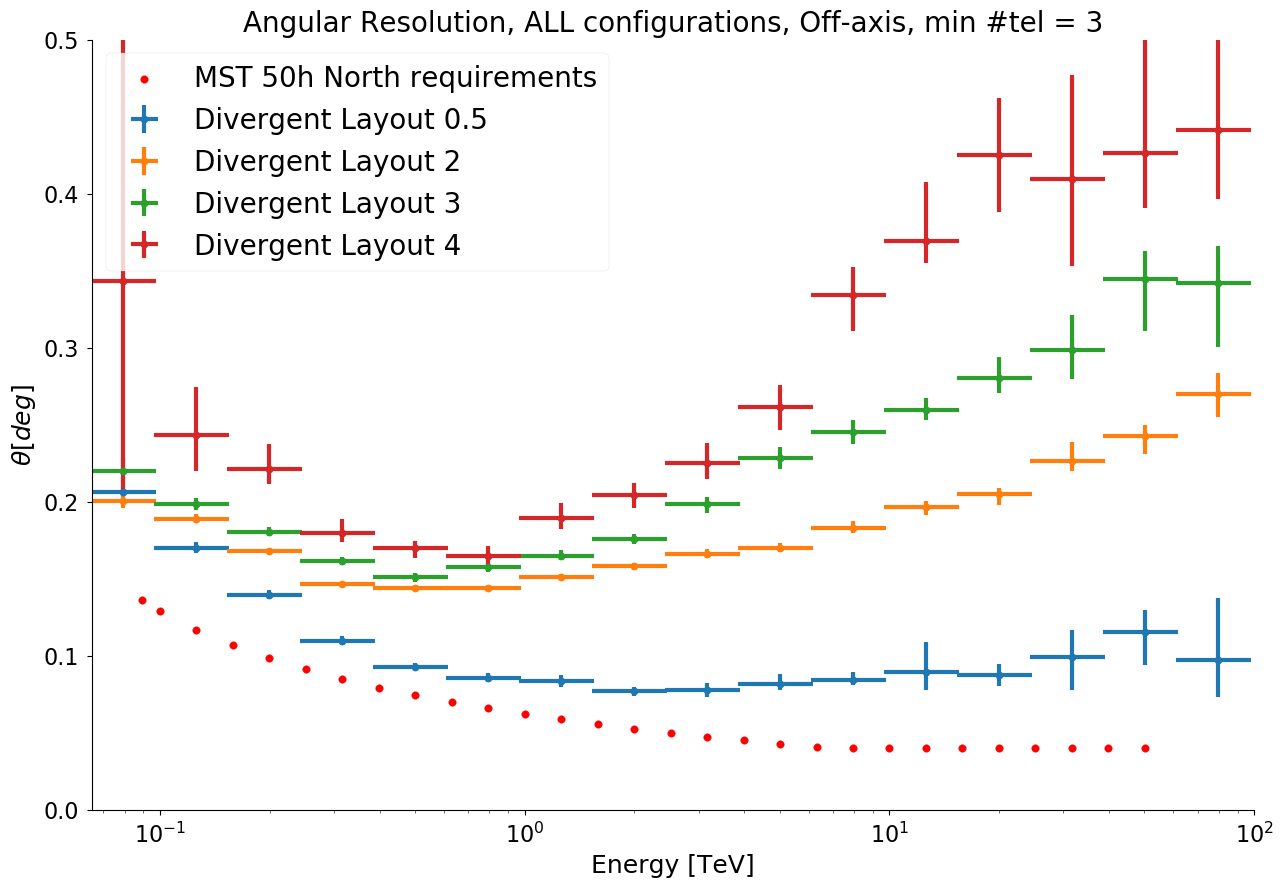}
		    \caption{Minimum multiplicity equal to 3.}
        \end{subfigure}
        \begin{subfigure}{\textwidth}
            \vspace{3mm}
            \centering
			\includegraphics[width=0.5\linewidth, keepaspectratio]{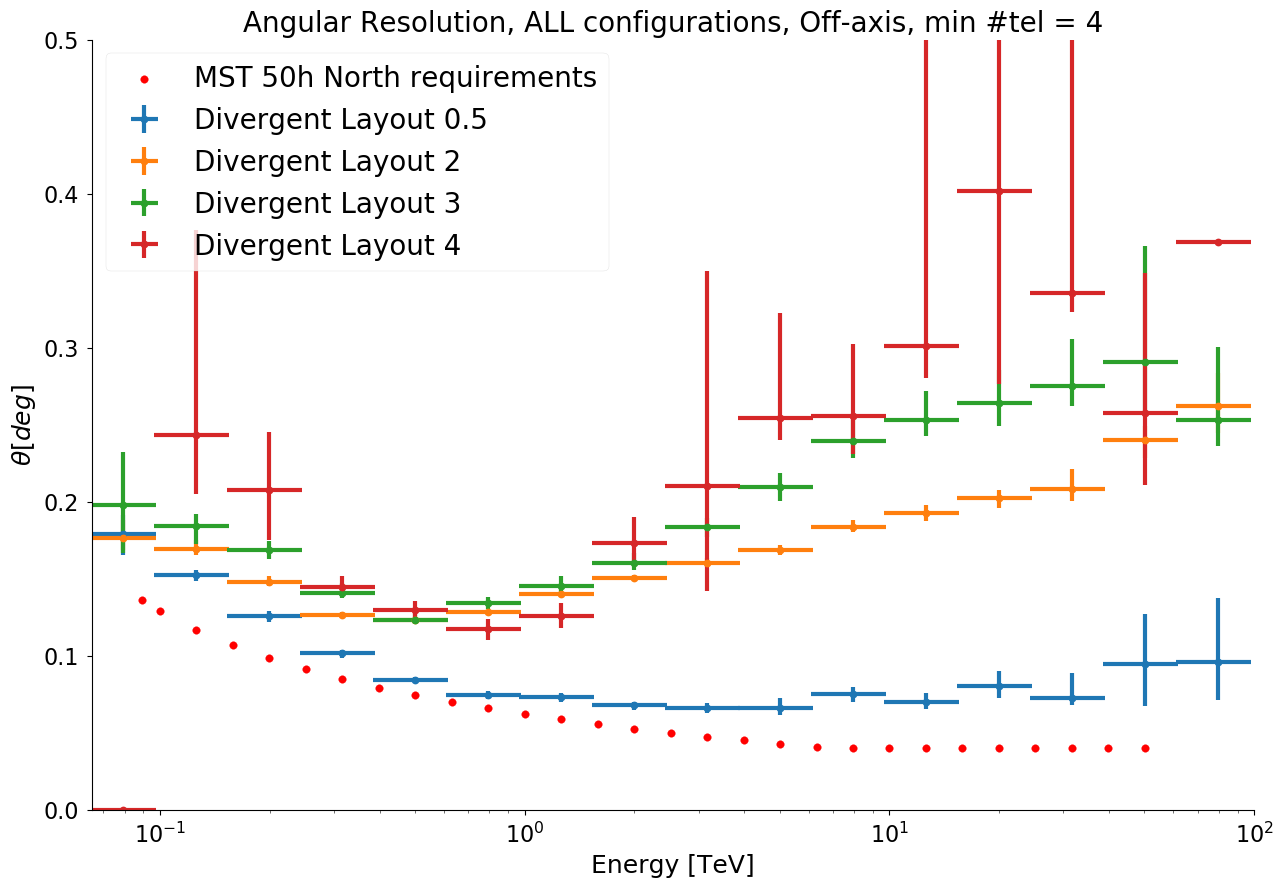}
		    \caption{Minimum multiplicity equal to 4.}
        \end{subfigure}
    \caption{Angular resolution of the "off-axis" configuration for the gamma-rays analysed plotted against the true energy. The plots (a), (b) and (c) have been created selecting only events reconstructed with at least 2, 3 and 4 telescopes triggered. Maximum leakage set to 10\%.}
    \label{Fig:AngResOffAxis}
    \end{figure}

    Together with the configurations shown in figure~\ref{Fig:fovs}, in which the average pointing of the telescopes is the position from which the gamma-rays are coming, we have also tested a similar set of configurations with the telescopes having a bigger offset with respect to the position of the gamma. In this second set of configurations, the telescopes' pointings are spread around an azimuth of 174 deg and an altitude of 70 deg (same pattern as plotted in figure~\ref{Fig:fovs}), with the gamma-rays coming from 180 deg of azimuth and 70 of altitude (results in figure~\ref{Fig:AngResOffAxis}).
    

\section{Conclusions}
In this contribution we have presented a status of the divergent pointing task for CTA and showed the first preliminary results obtained for different test configurations with a subarray of 15 MSTs using the configuration parameters of the latest MC production. We have adapted the reconstruction method for the parallel mode to be used also for the event reconstruction in divergent pointing.
Since we did not analyse any protons but only point source gamma-rays, as a next step we will produce angular resolution plots for events surviving the background suppression cuts, together with sensitivity plots.

An optimization and deeper study of the configurations of the telescopes pointing is necessary in the near future with the main goal to evaluate the performance of CTA operated in divergent mode.


\end{document}